\documentclass[aps,prb,twocolumn,groupedaddress]{revtex4}

\usepackage{graphicx}

\begin{document}

\newcommand{\BiM}{Bi$_2$$_-$$_x$M$_x$Ru$_2$O$_7$}
\newcommand{\Bi}{Bi$_2$Ru$_2$O$_7$}
\newcommand{\BiCo}{Bi$_2$$_-$$_x$Co$_x$Ru$_2$O$_7$}
\newcommand{\dg}{$^\circ$}
\newcommand{\BiCu}{Bi$_2$$_-$$_x$Cu$_x$Ru$_2$O$_7$}
\newcommand{\BiNi}{Bi$_2$$_-$$_x$Ni$_x$Ru$_2$O$_7$}
\newcommand{\BiMn}{Bi$_{1.9}$Mn$_{0.1}$Ru$_2$O$_7$}
\newcommand{\BiFe}{Bi$_{1.9}$Fe$_{0.1}$Ru$_2$O$_7$}

\title{Robust paramagnetism in    \BiM\ (M = Mn,Fe,Co,Ni,Cu) pyrochlore}

\author{M.K. Haas}
\author{R.J. Cava}
\affiliation{Department of Chemistry and Princeton Materials
Institute, Princeton University, Princeton, New Jersey 08540, USA}

\author{M. Avdeev}
\author{J.D. Jorgensen}
\affiliation{Argonne National Lab, Division of Materials Science,
Argonne, IL 60439, USA}

\date{\today}

%\preprint{preprint1}

\begin{abstract}
We report physical property characterization of \BiM\ pyrochlores,
including magnetic suseptibility, resistivity, and Seebeck
coefficients.  The solid solution exists up to x=0.5 for
(M=Cu,Ni,Co) and up to x=0.1 for (M=Fe,Mn). None of the doped
materials exhibit ferromagnetism or any localized ruthenium moment
behavior. Instead we find the Ru-O and Bi-O sublattices to be
essentially independent, with any magnetism resulting from the
unpaired transition metal dopant spins.  Cobalt substitution for
bismuth results in localized Co$^{2+}$, and low temperature
spin-glass transitions in several cases.  Nickel moments on the
pyrochlore lattice display properties intermediate to localized
and itinerant. Finally, copper doping results in only an
enhancement of the Pauli metallic density of states.
\end{abstract}


\maketitle

%%%%%%%%%%%%%%%%%%%%%%%%%%%%%%%%%%%%%%%%%%%%%%%%%%%%%%%%%%%%%% Introduction
\section{Introduction}
Ruthenium oxide based perovskites are presently of great interest
due to the range of magnetic and strongly correlated electronic
behavior they exhibit, often breaching the limits of current
condensed matter theory. Layered Sr$_2$RuO$_4$, for example, is a
superconductor at $<$ 1 K,\cite{maeno94} and is postulated to have
{\textit p}-wave and spin-triplet pairing in the superconduting
state.\cite{maeno01} Of particular interest is the delicate
balance many ruthenate perovskites display between ferromagnetism
and exotic electronic states. When superconducting Sr$_2$RuO$_4$
is doped by small amounts of nonmagnetic Ti$^{4+}$ on the
ruthenium site, there is a cross-over from superconductivity to
short range ferromagnetic ordering.\cite{minakata01, pucher02}  In
the same structural system, Sr$_3$Ru$_2$O$_7$ shifts from a
paramagnetic, strongly correlated Fermi-liquid to a ferromagnet
under applied pressure.\cite{ikeda01}  Also, at fields above 7T
the compound undergoes a metamagnetic quantum transition to a high
moment state.\cite{grigera01} Finally, Ti$^{4+}$ doping on the
ruthenium site of CaRuO$_3$ induces ferromagnetism in what was
originally a paramagnetic material.\cite{he1}

As much of the recent work focuses on perovskite based structures,
the question follows as to whether the observed balance between
itinerant and ferromagnetic behavior is characteristic of
ruthenium oxides in general, or specific to perovskites. Therefore
we have investigated another common oxide structure, the
pyrochlore. The pyrochlore oxide \Bi, which exhibits temperature
independent Pauli paramagnetism and is weakly metallic, affords an
excellent opportunity to test the generality of this balance.

\begin{figure}
\includegraphics[width=3in]{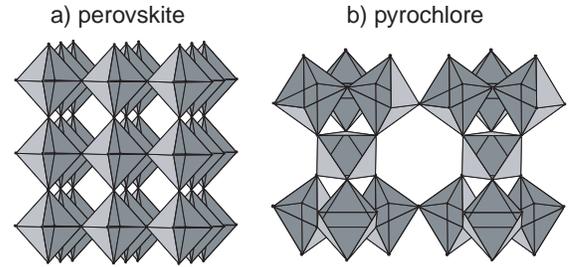}
\caption{ \label{structure} View of RuO$_6$ network in a)
perovskite along $<$100$>$ and b) pyrochlore along $<$011$>$.}
\end{figure}

Like the perovskite, the pyrochlore structure, A$_2$M$_2$O$_7$,
contains a three dimensional network of corner-sharing MO$_6$
octahedra (fig. \ref{structure}). In the perovskite these are
regular MO$_6$ octahedra with 160-180\dg\ M-O-M bonds, forming a
three dimensional square net of metal atoms. However, in the
pyrochlore the corner-sharing results in tetrahedrally related
metal atoms, and coordination is such that the M-O-M bond angle
must be $\sim$ 130\dg. The A cation in the pyrochlore occupies
hexagonal channels along $<$011$>$ and equivalent directions.
Three dimensionally, the cations form an A$_4$O sublattice of
corner sharing tetrahedra.

%                                                            %fig1  structure

Previous experimental work has probed differences between \Bi,
Y$_2$Ru$_2$O$_7$, and Ln$_2$Ru$_2$O$_7$, the latter two being
electronically insulating although isoelectronic to
\Bi.\cite{cox92}  One proposed reason for the difference is the
change in Ru-O-Ru angle with A cation size.\cite{kennedy98,lee97}
It is also postulated that in the compounds Y$_2$Ru$_2$O$_7$ and
Ln$_2$Ru$_2$O$_7$ the cation valence orbitals are energetically
prohibited from contributing to Fermi level bonding, resulting in
narrower localized bands.\cite{cox86} Reports are conflicting as
to whether bismuth \textsl{s} and \textsl{p} orbitals hybridize to
some extent at the Fermi level, participating in
conduction.\cite{ishii00,hsu88} Experimental work has followed the
metal to insulator transition in several of these solid
solutions.\cite{yoshii00,yamamoto94} Here we report the properties
of \BiM, M =(Mn,Fe,Co,Ni,Cu), observing that the first row
transition metal d-orbitals are proximate to the energy of the
Fermi level of \Bi. Thus it is likely that the overall properties
of these doped materials will be markedly different from those
previously described in the literature.

%%%%%%%%%%%%%%%%%%%%%%%%%%%%%%%%%%%%%%%%%%%%%%%%%%%%%%%%  Experimental
\section{Experimental}
Suitable stoichiometric amounts of RuO$_2$, Bi$_2$O$_3$,
Co$_3$O$_4$, CuO, MnO$_2$, Fe$_2$O$_3$, and Ni$_2$O$_3$ were
thoroughly ground in an agate mortar and pressed into pellets.
RuO$_2$ was dried for 2 hours at 900 \dg C before use.  The
pellets were successively heated in air at 750 \dg C for 1 day,
900 $^\circ$C for 1 day, 950 \dg C for 3 days, and 975 \dg C for 1
day, with several intermediate grindings.  Iron and nickel samples
were heated for an additional day at 1020 \dg C.  Sample purity
was monitored using room temperature powder X-ray diffraction
employing Cu-K$\alpha$ radiation.

Magnetic properties and electrical resistivity were evaluated
using a Quantum Design physical property measurement system
(PPMS).  For all samples, the suseptibilty was measured from 300
to 5 K in an applied field of 1 T.  Data were also collected from
2 to 30 K in a field of 1 T after zero-field cooling, to assess
any low temperature magnetic transitions. Magnetic hysteresis
loops were performed at 5 K in the range of -9 to 9 T, and the
data were fit to the paramagnetic Brillouin function $M = NgJ\mu
_B B(x)$ where $x = m\mu_0H/k_BT$.\cite{brillion27} Resistivity
measurements were performed on sintered powder pellets, which were
cut into bars of approximately 1.5$\times$1.5$\times$3.5 mm, and
measured with the standard 4-point AC method. Seebeck coefficient
measurements employed a commercial apparatus (MMR technologies)
within the temperature range 200-500 K.

%%%%%%%%%%%%%%%%%%%%%%%%%%%%%%%%%%%%%%%%%%%%%%%%%%%%%%%%%%%% Results
\section{Results}
Substitution of first row transition metals on the ruthenium site
of \Bi\ was not possible within the synthetic conditions explored.
The first row transition metals Mn, Fe, Co, Ni, and Cu instead
substitute on the bismuth site of the pyrochlore. We report
complete structural characterization of
Bi$_{1.6}$Cu$_{0.4}$Ru$_2$O$_7$, Bi$_{1.6}$Co$_{0.4}$Ru$_2$O$_7$,
and \Bi\ elsewhere,\cite{avdeev02} by refinement of neutron powder
diffraction data. This type of substitution is not unprecedented,
as synthesis of Bi$_{2-x}$Cu$_x$Ir$_{2-y}$Ru$_y$O$_7$ was
reported, although the compound was not fully
characterized.\cite{beck86}

%                                                     %fig2 lattice
\begin{figure}
\includegraphics[width=3in]{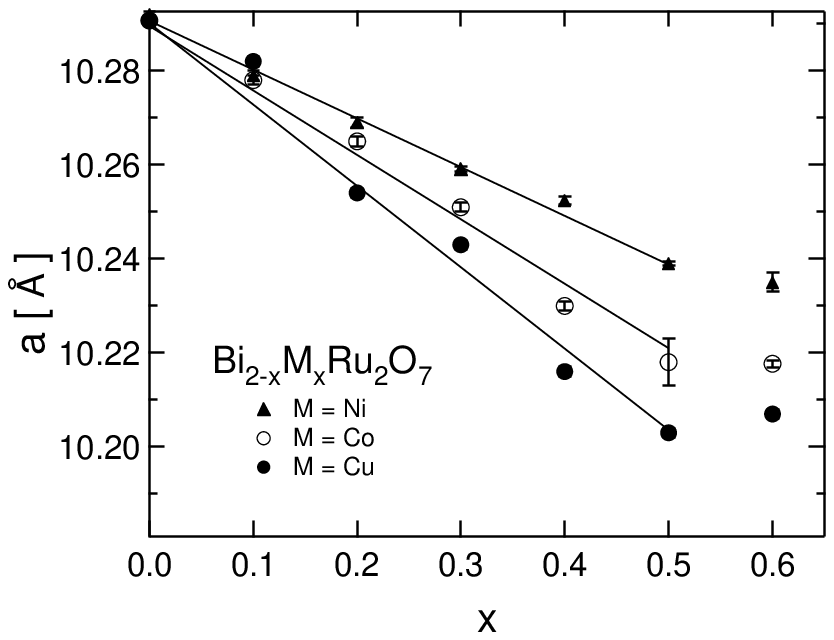}
\caption{\label{lattice}Cubic lattice parameter,determined by
powder x-ray diffraction, for \BiM\ (M=Cu,Ni,Co) as a function of
doping level.}
\end{figure}

 Figure \ref{lattice} plots lattice parameters obtained from powder
X-ray diffraction as a function of x for \BiM\ (M=Cu,Ni,Co).  To a
first approximation the lattice parameters decrease linearly. This
is consistent with the smaller radii of the first row transition
metals in comparison to that of bismuth.  For copper, nickel, and
cobalt dopants, the solid solution exists up to x=0.5: at larger
nominal concentrations the lattice parameters remain relatively
constant. In several samples, RuO$_2$ is found as a very minor
impurity.  For example, the cobalt-doped sample with x=0.4,
contains 0.5\% RuO$_2$ by weight, as determined by powder neutron
diffraction refinement.\cite{avdeev02}  As RuO$_2$ is
non-magnetic, and the amount present is negligible, its presence
will not affect physical property characterization. For iron and
manganese-doped series the substitution limit is x=0.1, resulting
in unit cells of a=10.282(1) and 10.279(2) respectively.

\subsection{\BiCu}

\begin{figure}
\includegraphics[width=3in]{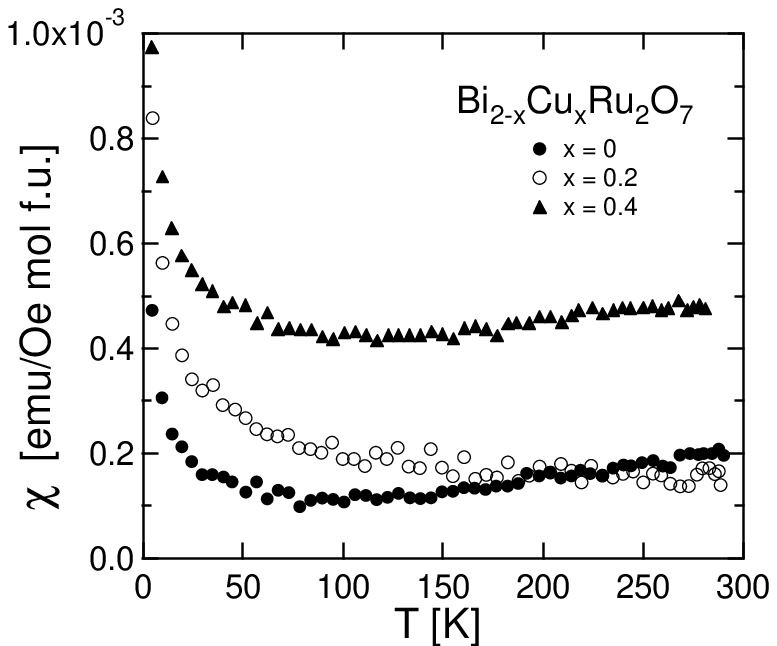}
\caption{\label{chi_cu} Measured magnetic suseptibility as a
function of temperature for \BiCu.}
\end{figure}

The magnetic suseptibility ($\chi$) of \Bi\ between 300 and 5 K is
plotted in figure \ref{chi_cu} .  It is essentially temperature
%                                                     %figure3  chi_cu
independent with a relatively small magnitude of 2 $\times$
10$^{-4}$ emu/Oe mol f.u., consistent with Pauli paramagnetism.
Also plotted in figure \ref{chi_cu} is the suseptibility of
Bi$_{1.6}$Cu$_{0.4}$Ru$_2$O$_7$ and
Bi$_{1.8}$Cu$_{0.2}$Ru$_2$O$_7$.  It can be seen that x=0.4
copper-doping increases the magnitude of the temperature
independent suseptibility by a factor of two. This may be due to
an enhancement of the metallic density of states. There is not any
significant indication of moment localization for either copper or
ruthenium in these samples. Each does display a small Curie tail
at low temperature, likely due to the presence of microscopic
amounts of impurity spins.

\begin{figure}
\includegraphics[width=3in]{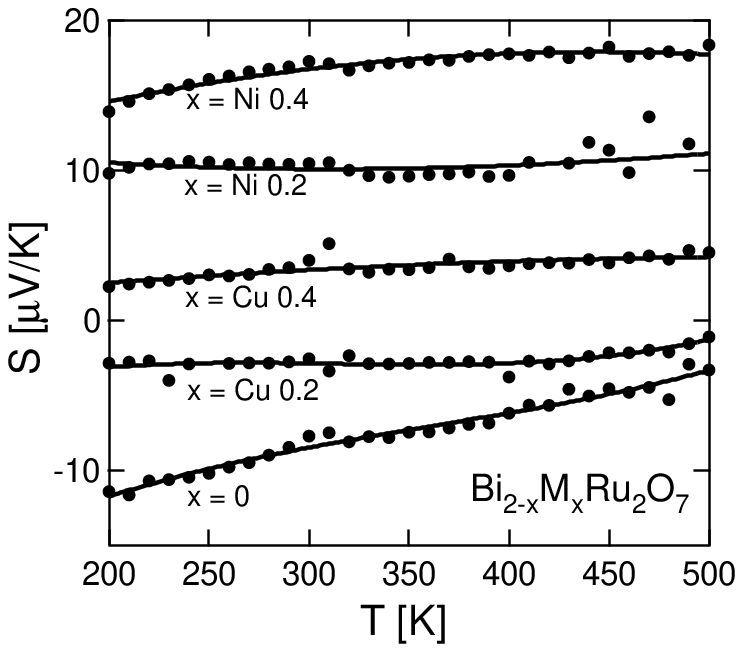}
\caption{\label{s_cu}Seebeck coefficients for \BiM\ (M=Cu,Ni) as a
function of temperature.}
\end{figure}

Seebeck coefficients (S) for \BiCu\ are plotted in figure
\ref{s_cu}. (Small peaks evident in the data around 300 K are
%                                                         %fig4  s_cu
instrumental.) It can be seen that values of S for undoped \Bi\
are between -11 and -4 $\mu$V/K, indicating that the dominant
carriers are electrons. Ten percent copper doping (x=0.2)
increases S, though the values remain negative, between -3 and -1
$\mu$V/K. However with twenty percent (x=0.4) doping, the seebeck
coefficients become positive, and the dominant carriers cross over
to hole-like.

\begin{figure}
\includegraphics[width=3in]{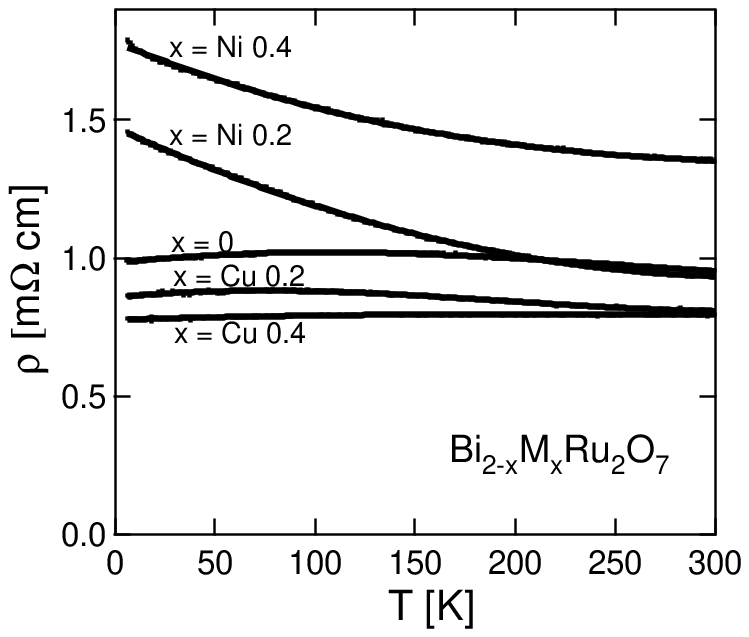}
\caption{\label{r_cu}Resistivity as a function of temperature for
\BiM\ (M=Cu,Ni).}
\end{figure}

Resistivity data are presented in figure \ref{r_cu}. The
resistivity of undoped \Bi\ is largely temperature independent and
on the order of 1 m$\Omega\cdot$cm, consistent with the
description of a poor metal. Copper doping slightly decreases the
magnitude of the resistivity but maintains a similar temperature
dependence compared to the undoped sample.
                                                            %%fig5 r_cu

\subsection{\BiNi}

\begin{figure}
\includegraphics[width=3in]{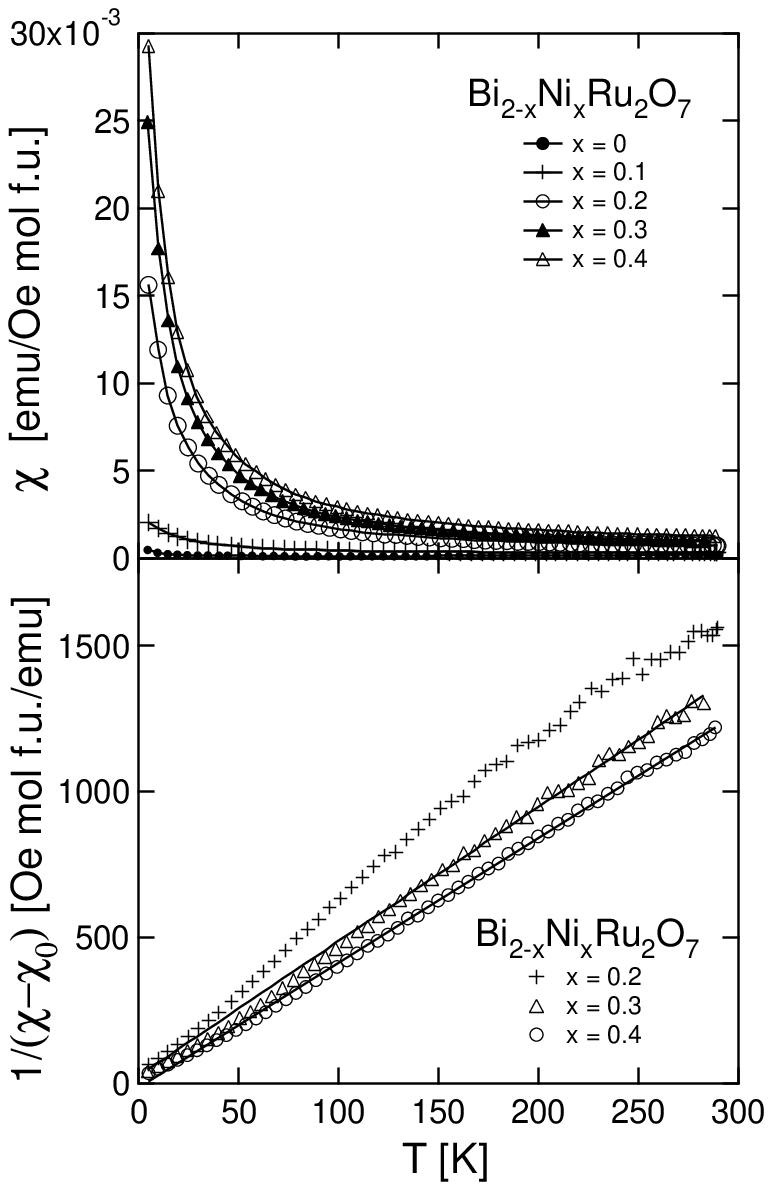}
\caption{\label{chi_ni}Top panel:  Measured magnetic suseptibility
as a function of temperature for \BiNi.  Bottom panel:  Inverse
magnetic suseptiblity (points) and Curie-Weiss fits (solid lines)
for \BiNi.  Fitted temperature independent terms ($\chi_0$) are
subtracted from measured $\chi$.}
\end{figure}

Magnetic suseptibility data for nickel-doped samples are shown in
figure \ref{chi_ni}.  The top panel plots measured suseptibility
as a function of temperature.  In contrast to copper doping, these
%
                                                            %fig6 chi_ni
samples begin to display local moment behavior as nickel is
substituted. The Curie-tail at low temperature that is visible on
the undoped pyrochlore becomes more pronounced with 5\% (x=0.1)
nickel doping. By 20\% (x=0.4) nickel doping, the magnetic
suseptibility scales with the Curie-Weiss equation for localized
paramagnetic moments. Data from nickel-doped samples for x=0.3 and
x=0.4 were fit between 150-300 K to the function $\chi = \chi_0 +
C/(T-\theta_{CW})$ where $\chi_0$ is the sum of all temperature
independent terms.  Values for the effective moment and
Curie-Weiss temperature ($\theta_{CW}$) were extracted, with the
results summarized in table \ref{fits}. Moments are between 2.1
and 2.3 $\mu_B$/Ni, only slightly less than the theoretical
spin-only value of 2.82 for Ni$^{2+}$. Therefore we assume that
the observed local moments are accounted for by the nickel dopant,
and that no local moment has been induced at ruthenium centers.
Further, values of $\theta_{CW}$ are very small, indicating weak
coupling between the localized spins in all cases.

\begin{table}
\caption{\label{fits} Calculated magnetic constants from
Curie-Weiss fitting.}
\begin{tabular}{|c|c|c|c|} \hline
  Dopant x & {$\chi_0$} & {$\bf \theta_{CW}$} & {\bf
      $\mu/\mu_B$} \\
  {} & [emu/Oe mol f.u.] & {[K]} & [per x] \\ \hline
  Ni 0.3 & 2.3 $\times$ 10$^{-4}$ & -4.0(9) & 2.39(1) \\
  Ni 0.4 & 4.4 $\times$ 10$^{-4}$ & \ 3.7(6) & 2.161(3) \\
  Co 0.1 & 2.2 $\times$ 10$^{-4}$ & \ 1.3(3) & 4.381(4) \\
  Co 0.2 & 3.8 $\times$ 10$^{-4}$ & \ 2.1(3) & 4.265(4) \\
  Co 0.3 & 3.8 $\times$ 10$^{-4}$ & \ 1.9(5) & 4.129(4) \\
  Co 0.4 & 2.5 $\times$ 10$^{-4}$ & \ 2.1(2) & 4.465(2) \\
  Co 0.5 & 3.0 $\times$ 10$^{-4}$ & -0.7(4) & 4.269(2) \\
  Fe 0.1 & 3.3 $\times$ 10$^{-4}$ & \ 0.1(3) & 4.937(5) \\
  Mn 0.1 & 1.6 $\times$ 10$^{-4}$ & \ 1.5(4) & 5.112(6) \\ \hline
\end{tabular}
\end{table}

The bottom panel of figure \ref{chi_ni} plots inverse magnetic
suseptibility, 1/($\chi - \chi_0$), for the nickel-doped samples.
Experimental points are overlayed by Curie-Weiss fits (solid
lines) for x=0.3 and x=0.4.  The overlays highlight evident
deviation from ideal behavior below 100 K for the two samples. And
in fact, at nickel doping levels of x=0.2 and lower, deviation
from linear behavior is such that Curie-Weiss fitting is not
applicable.  Thus the magnetism can be classified as intermediate
to localized and itinerant behavior.

Seebeck coefficients (S) for \BiNi\ are presented in figure
\ref{s_cu}. At 200 K, the magnitude of the coefficient goes from
-10 in undoped \Bi\, to +10 $\mu$V/K in the 10$\%$ nickel-doped
sample. The nickel-doped samples have higher S values than the
copper-doped samples, indicating that there is a higher
concentration of hole carriers in the former.  Resistivity
($\rho$) behavior is also different for the copper and
nickel-doped samples.  Figure \ref{r_cu} illustrates that while
copper-doping decreases the magnitude of the resistivity,
nickel-doping increases it.  As nickel is substituted on the
lattice of \Bi\, the resistivity becomes more like that of a
degenerate semi-conductor, where $\rho$ begins to increase
slightly with decreasing temperature.

\subsection{\BiCo}
\begin{figure}                                       %figure7 chi_co
\includegraphics[width=3in]{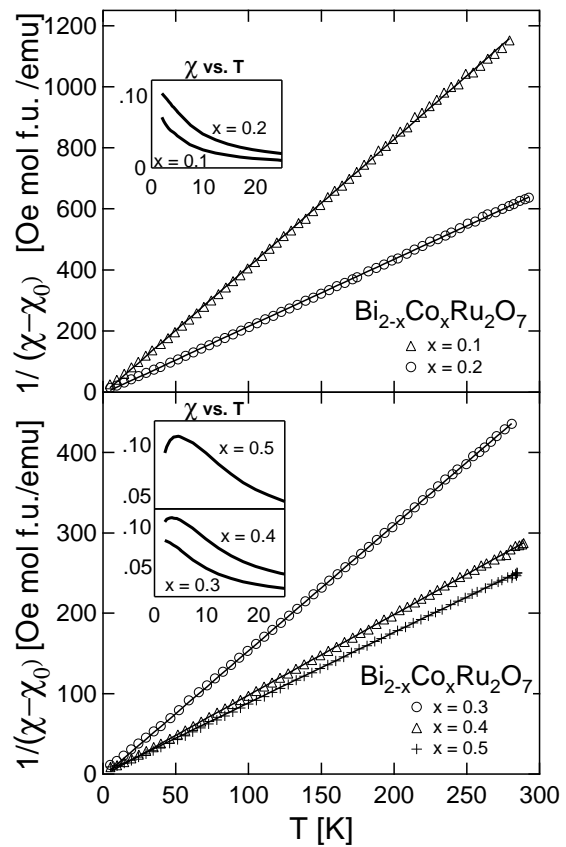}
\caption{\label{chi_co}Inverse magnetic suseptibility (points) and
Curie-Weiss fits (solid lines) for \BiCo\ as a function of
temperature.  Fitted temperature independent terms ($\chi_0$) are
substracted from measured $\chi$. Insets plot low temperature
magnetic suseptibility down to 2 K.}
\end{figure}

Cobalt doping has a dramatic effect on the magnetic properties of
\Bi. Samples of \BiCo\ display Curie-Weiss behavior for x=0.1 to
0.5 in the temperature range 300-5 K and in 1T.    Plots of
1/$(\chi - \chi_0)$ versus temperature (fig. \ref{chi_co})
demonstrate the
goodness of fit, where calculated values (solid lines) overlay
experimental points.  The effective moment normalized to cobalt
content (table \ref{fits}) remains essentially constant across the
doping series, indicating that the Curie moments are due solely to
unpaired cobalt d electrons. Again no local moment is induced on
ruthenium centers with doping. Calculated moments are between 4.1
and 4.4 $\mu_B$/Co, which are slightly greater than the spin-only
theoretical value of 3.87 for Co$^{2+}$ but consistent with
previously reported values for high spin Co$^{2+}$ of 4.1 to 5.2
$\mu_B$/Co.\cite{berger73} Values of theta are close to zero for
all cobalt doped samples, signifying that spin-coupling is very
weak. Below 5 \dg K, small peaks in magnetic suseptibility are
observed for x=0.4 and x=0.5 samples, possibily due to spin-glass
transitions (fig. \ref{chi_co}, inset).  The fact that the
transitions are at such low temperatures is consistent with weak
coupling of the spins. Additionally, the peak maximum increases
slightly from samples x=0.4 to x=0.5. In fact, although there is
not a distinct suseptibility peak for sample x=0.3, there are
significant deviations from Curie-Weiss behavior below 5 K,
signifying that some degree of interaction is present already at
this doping level.  The doped pyrochlores have the two
characteristics common to spin-glass systems; atomic disorder and
frustrated geometry.

\begin{figure}
\includegraphics[width=3in]{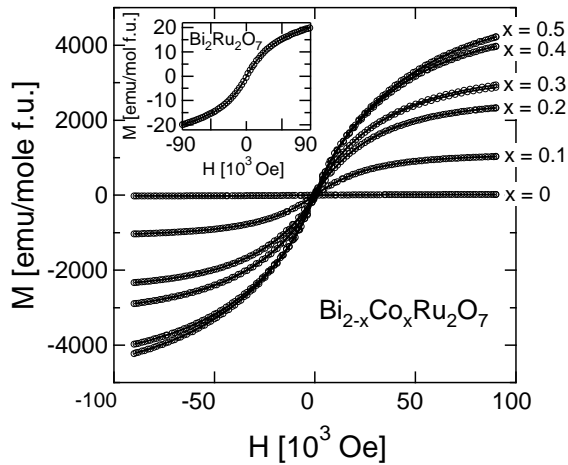}
\caption{\label{loop_co}Magnetization as a function of field for
\BiCo. Inset: Expansion of measured magnetization versus field
data for undoped \Bi\ in low temperature region. Units are: $\chi$
(emu/Oe mol f.u.), T (K). }
\end{figure}

\begin{figure}
\includegraphics[width=3in]{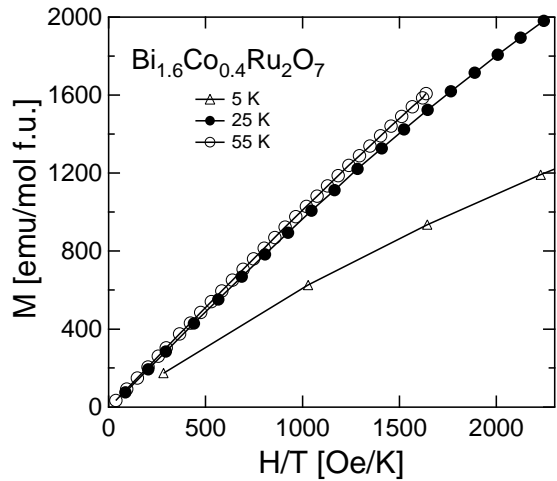}
\caption {\label{ht_co}Magnetization as a function of H/T for
Bi$_{1.6}$Co$_{0.4}$Ru$_2$O$_7$, with data sets collected at three
different temperatures.}
\end{figure}

Magnetization (M) versus field (H) curves are presented in figure
\ref{loop_co} for \BiCo, where data are collected at 5 K.  No
magnetic hysteresis is present for any doping level.   The curves
are not
%                                                        %fig8 loop_co
linear, but instead scale with the paramagnetic Brillion function,
as m$\mu_B$H is on the order of k$_B$T at this temperature and
field range.  Although M vs H for undoped \Bi\ does appear to be
flat and linear at this scale, the figure inset shows the true
shape of the curve.  Fitting the high field data for undoped \Bi\
to the Brillouin function with g=2, the number of localized spin 1
moments is approximated to be 0.002 spins/mol f.u.  For \BiCo\,
the magnetization increases with the amount of cobalt dopant as is
expected.  However, attempts to fit these curves to the Brillion
function were unsuccessful.  This is due to the fact that the data
were collected at 5 K, where an assumption of completely localized
non-interacting moments is inappropriate for these samples. Figure
\ref{ht_co} plots magnetization as a function of H/T for
%                                                            %fig9 ht_co
sample Bi$_{1.6}$Co$_{0.4}$Ru$_2$O$_7$ at different temperatures.
If the data follow the paramagnetic Brillouin function, then plots
taken at different temperatures should superimpose.  In figure 9
it is evident that plots for 25 K and 55 K data essentially
overlay one another. However, data taken at 5 K deviates
significantly. Therefore the cobalt moments are truly independent
at higher temperatures, but by 5 K, some degree of interaction is
present between the cobalt centers. This is supported by our
suseptibility data, where possible spin-glass ordering is present
below 5 K for the more heavily doped samples.  It is surprising
that such a large number of magnetic Co$^{2+}$ atoms are
accommodated in the pyrochlore without inducing magnetism on the
ruthenium sublattice, and that the Co$^{2+}$ spins are so weakly
interacting.

\begin{figure}
\includegraphics[width=3in]{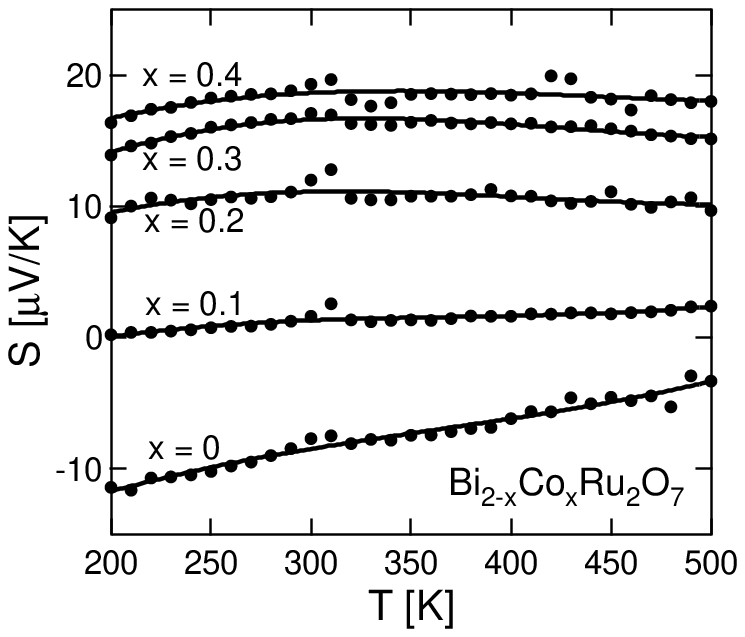}
\caption{\label{s_co}Seebeck coefficients as a function of
temperature for \BiCo.}
\end{figure}

\begin{figure}
\includegraphics[width=3in]{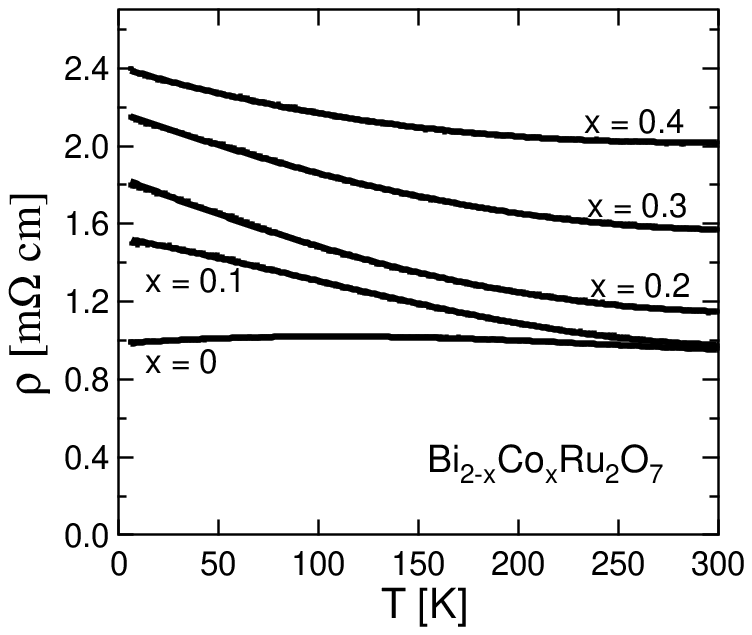}
\caption{\label{r_co}Resistivity as a function of temperature for
\BiCo.}
\end{figure}

\begin{figure}
\includegraphics[width=3in]{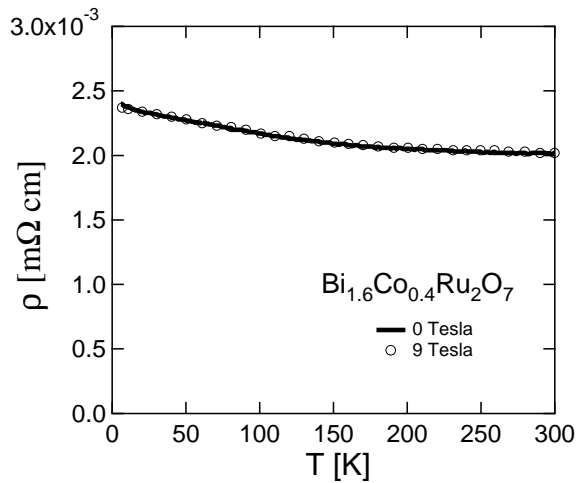}
\caption{\label{cmr_co}Resistivity as a function of temperature
for Bi$_{1.6}$Co$_{0.4}$Ru$_2$O$_7$ at 0 Tesla and 9 Tesla.}
\end{figure}

Seebeck coefficients and resistivity for \BiCo\ are plotted in
figures \ref{s_co} and \ref{r_co} respectively.  Seebeck
coefficients increase
%
                                                     %fig10 s_co
%
%                                                    %fig11 r_co
with the amount of cobalt doping, reaching a maximum of +18
$\mu$V/K for x=0.4.  Once again, the first row transition metal is
hole-doping the pyrochlore.  The slope of the temperature
dependence of S also becomes flatter with increasing cobalt. The
resistivity behavior of \BiCo\ is similar to that of \BiNi.  As
dopant is introduced, the magnitude of the resistivity increases,
and the temperature dependence becomes more semi-conductor like.
The increased resistivity may be due to several factors, one of
which is the scattering of electrons by the localized dopant
moments. This explanation is supported by the fact that in
copper-doped samples, where there are no localized spins, the
resistivity decreases slightly, instead of increasing.  However,
the resisitivity of Bi$_{1.6}$Co$_{0.4}$Ru$_2$O$_7$ is unaffected
by a field of 9T (figure \ref{cmr_co}).  This is not consistent
with the description of spin scattering, where at 9T the spins
should align and thus decrease the resistance.  Therefore the
origin of resistivity behavior in copper and nickel doped \Bi\ is
unresolved.

%                                                           %fig12 cmr_co

\subsection {\BiFe and \BiMn}

%                                                            %fig13 chi_mn
\begin{figure}
\includegraphics[width=3in]{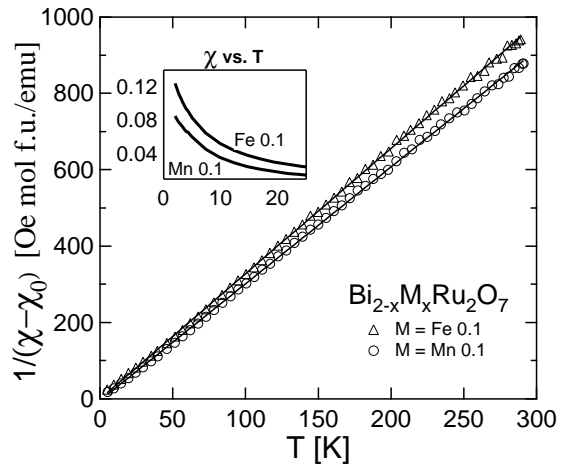}
\caption{\label{chi_mn}Inverse suseptibility (points) and
Curie-Weiss fits (solid lines) for \BiMn\ and \BiFe\ as a function
of temperature. Fitted temperature independent terms ($\chi_0$)
are subtracted from measured $\chi$. Insets plot measured magnetic
suseptibility down to 2 K. Units are: $\chi$ (emu/Oe mol f.u.), T
(K). }
\end{figure}

\begin{figure}
\includegraphics[width=3in]{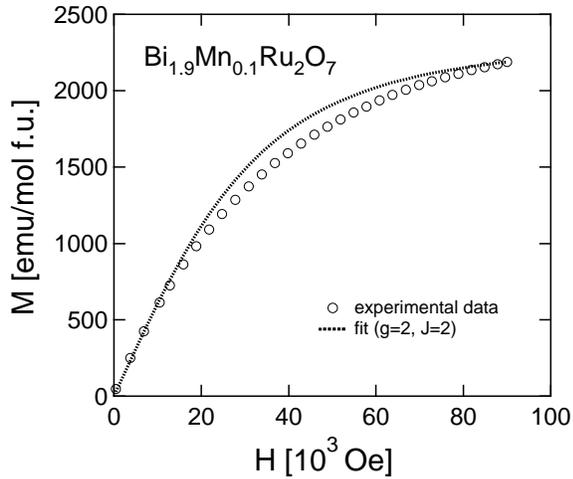}
\caption{\label{b_mn}Experimental magnetization vs. field data for
\BiMn\ and paramagnetic Brillouin function (dotted line).}
\end{figure}

For dopants Fe and Mn, the substitution limit is 5$\%$ (x=0.1).
Inverse suseptibility for these two samples is plotted in figure
\ref{chi_mn}, with overlayed Curie-Weiss fits (solid lines).  Both
\BiFe\ and \BiMn\ obey the Curie-Weiss law in the temperature
range 300-5 K.  Magnetic constants are listed in table \ref{fits}.
Fitted magnetic moments are 4.9 and 5.1 $\mu_B$/x for Fe and Mn
respectively. These values are in good agreement with the
spin-only theoretical value of 4.89 $\mu_B$ for S=2, again
indicating that the observed localized moments are due to the
dopants.  Values of $\theta_{CW}$ are close to zero, indicating
that interaction between the magnetic centers are negligible. In
fact, the manganese spins are truly independent even down to low
temperature.  This is illustrated in figure \ref{b_mn}, where
magnetization is plotted as a function of field for \BiMn. The
high field data were fit to the paramagnetic Brillouin function
with set values g=2 and J=2, and with only N (number of spins per
mole) allowed to vary.  Assuming that all localized spins
originate from manganese, the resulting value of x found for
\BiMn\ was 0.10, in agreement with the nominal concentration.

%                                                         %fig14 b_mn

\begin{figure}
\includegraphics[width=3in]{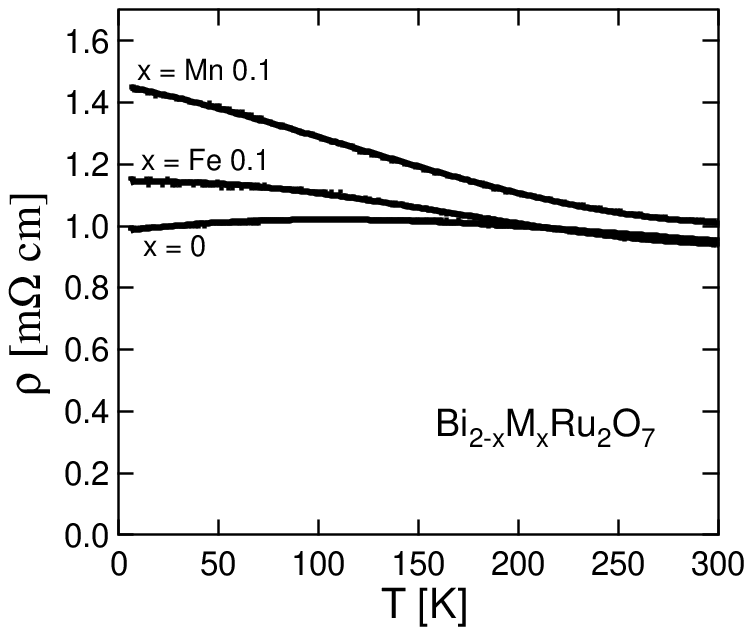}
\caption{\label{r_mn}Resistivity as a function of temperature for
\BiM\ (M=Fe,Mn).}
\end{figure}

\begin{figure}
\includegraphics[width=3in]{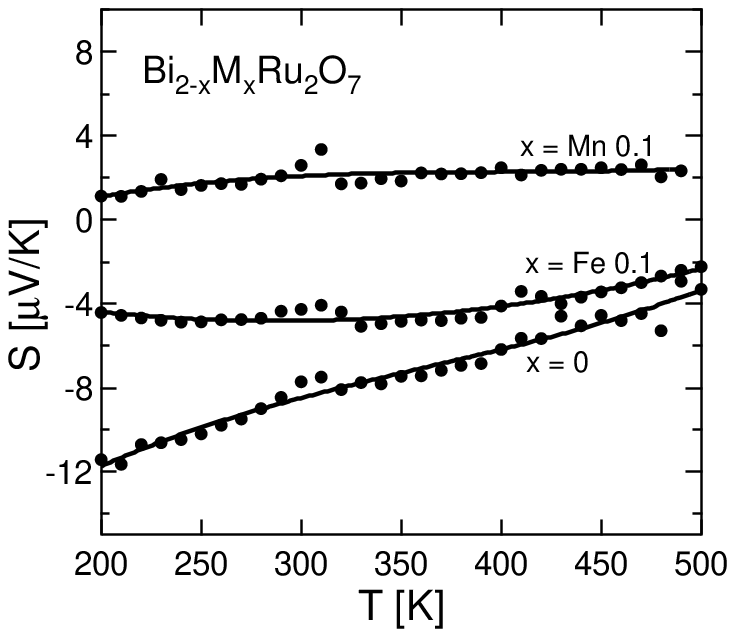}
\caption{\label{s_mn}Seebeck coefficient as a function of
temperature for \BiM\ (M=Fe,Mn).}
\end{figure}

 Resistivity (fig. \ref{r_mn})and Seebeck coefficients (fig. \ref{s_mn})
for Mn and Fe-doped samples follow similar trends as those seen
for Ni and Co.  The introduction of dopant increases the magnitude
of the resistivity, and samples become more semi-conducting.  For
Fe and Mn-doping, Seebeck coefficients also increase relative to
undoped \Bi\, indicating that the transition metals are once again
hole doping the pyrochlore.

%                                                             %fig15 r_mn

%                                                              %fig16 s_mn

\section{Discussion}

The pyrochlore structure can be described as two interpenetrating
networks; one of corner-sharing RuO$_6$ polyhedra and one of
corner-sharing Bi$_4$O tetrahedra.  Our magnetic data on
transition metal doped \Bi\ indicates that the two networks are,
surprisingly, magnetically independent.

 We successfully substituted the first row transition metal series
Mn$^{3+}$, Fe$^{2+}$, Co$^{2+}$, Ni$^{2+}$, and Cu$^{2+}$ on the
bismuth site of \Bi.  Each of these dopants has unpaired valence
shell d-electrons. However, the ruthenium moments in the
pyrochlore are unaffected by Bi-site doping. This is in strong
contrast to results for the doped perovskites
CaRu$_{1-x}$Ti$_x$O$_3$ and Sr$_2$Ru$_{1-x}$Ti$_x$O$_4$. Although
we find no ruthenium local moments, the unpaired
%                                                                   %fig17 c1
\begin{figure}
\includegraphics[width=3in]{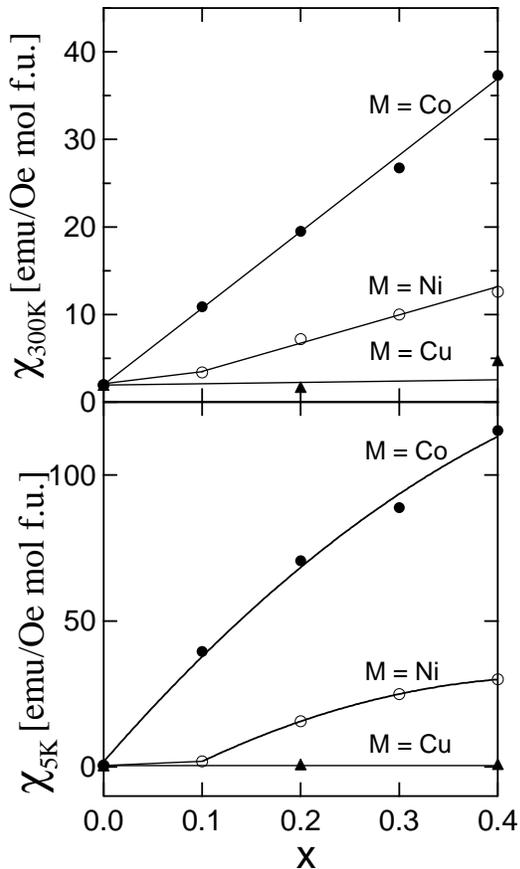}
\caption{\label{c1}Magnetic suseptbilibty at 300 K (top panel) and
5 K (bottom panel) as a function of dopant for copper, nickel, and
cobalt doped samples.}
\end{figure}
dopant electrons diverge from localized to itinerant magnetic
behavior across the first row transition metal series. First, in
\BiMn\, the substituted manganese moments are localized and
essentially noninteracting, even down to low temperature. In
\BiCo\  the cobalt moments are also localized, with suseptibility
data fitting the Curie-Weiss law between 5-300 K.  However, at low
temperature the spins are no longer completely isolated.  In fact,
spin-glass transitions are present below 5 K for
Bi$_{1.6}$Co$_{0.4}$Ru$_2$O$_7$ and
Bi$_{1.5}$Co$_{0.5}$Ru$_2$O$_7$.  In nickel-doped samples the
unpaired nickel spins display behavior intermediate to localized
and itinerant.  Finally, in copper-doped samples, there is no
evidence for moment localization, and only the metallic
temperature independent paramagnetism is enhanced with doping.
These magnetic trends are illustrated in figure \ref{c1}, where
the measured suseptibility at 5 K and 300 K is plotted as a
function of dopant.  For localized moments, the magnetic
suseptibility at low temperature increases observably as a
function of x, such as in the case of cobalt.  However, for
itinerant moments as in copper, $\chi_{5K}$ remains relatively
flat with increasing x. For nickel, with atomic number
intermediate to Co and Cu, the behavior is a combination of
localized and itinerant. Up to x=0.1 $\chi_{5K}$ remains flat,
while above x=0.1 $\chi_{5K}$ increases with x, although not as
steeply as $\chi_{5K}$ in the case of cobalt.  Across the
 series, the d-orbitals of the first row
transition metal become more proximate in energy to the valance
orbitals of oxygen. Thus copper hybridizes more strongly with
oxygen than cobalt or nickel, resulting in itinerant behavior.

                                                        %fig18 c2
\begin{figure}
\includegraphics[width=3in]{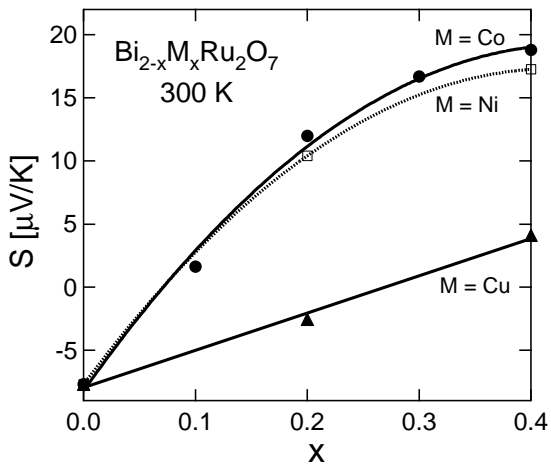}
\caption{\label{c2}Seebeck coefficients as a function of dopant
concentration for copper, nickel, and cobalt doped \Bi.}
\end{figure}

The transport properties of \BiM\ are also evaluated.  Figure
\ref{c2} displays S as a function of dopant concentration for Co,
Ni, and Cu doped samples.  At x=0.4, values of S are all positive.
Thus in each series the dominant carriers cross-over from
electrons to holes upon doping.  The absolute value of the Seebeck
coefficient, $\mid$S$\mid$, is proportional to log(N)-log(n) where
n is the concentration of carriers and N the available density of
states. If N is essentially constant, then copper doping results
in an increased concentration of carriers (n) relative to undoped
\Bi. This interpretation agrees with resistivity data, where the
magnitude of $\rho$ decreases as copper is substituted on the
pyrochlore lattice.  Conversely, Seebeck coefficients for cobalt
and nickel samples increase more steeply with x than those for
copper. At x=0.4 for these two dopants, the concentration of
carriers (n) is decreased relative to undoped \Bi.

\section{conclusion}
In summary, we report magnetic and transport properties of \BiM\
(M=Mn,Fe,Co,Ni,Cu).  In no case did ferromagnetism or other
localized ruthenium moment behavior occur upon doping.  However,
in such previously reported cases as CaRu$_{1-x}$Ti$_x$O$_3$ and
Sr$_2$Ru$_{1-x}$Ti$_x$O$_4$, the dopant was substituted on the
ruthenium site. Thus those substitutions could be expected to
strongly perturb the Ru-O network.  In the present study, the
transition metal dopants preferentially replace bismuth in \Bi.
Our magnetic data indicates that the ruthenium and bismuth
sublattices in \Bi\ are magnetically independent, and therefore
the metal substitutions represent a second-order perturbation of
the Ru-O lattice, which was not effective in causing a crossover
to the ferromagnetic state.  In CaRuO$_3$, conversely, A-site
substitution does induce ferromagnetism even though it is not a
direct perturbation of the Ru-O lattice.\cite{he2}  Therefore
ruthenium oxides in the pyrochlore geometry appear to be far from
the ferromagnetic instability observed in the perovskites, and
such instability is not a broadly general characteristic of
ruthenium oxide compounds.

\begin{acknowledgments}
This work was supported by the National Science Foundation Grant
No. DMR-9808941.  The work at Argonne National Laboratory was
supported by the US Department of Energy, Office of Basic Energy
Science, contract No. W-31-109-ENG-38. 
\end{acknowledgments}


\end{document}